\documentclass[11pt]{article}
\usepackage{latexsym}  
\usepackage{amssymb}
\usepackage{graphicx}
\usepackage{amsmath}

\begin{document}

\begin{center}

{\Large\bf Structure of Right-Handed Neutrino Mass Matrix}

\vspace{4mm}

{\bf Yoshio Koide}

 {\it Department of Physics, Osaka University, 
Toyonaka, Osaka 560-0043, Japan} \\
{\it E-mail address: koide@kuno-g.phys.sci.osaka-u.ac.jp}

\end{center}

\vspace{3mm}

\begin{abstract}
Recently, Nishiura and the author have proposed a unified 
quark-lepton mass matrix model under a family symmetry 
U(3)$\times$U(3)$'$.  The model can give excellent 
parameter-fitting to the observed quark 
and neutrino data.  The model has a reasonable basis 
as far as the quark sector, but the form of the right-handed 
neutrino mass matrix $M_R$ does not have a theoretical grand, 
that is, it was nothing but a phenomenological assumption.   
In this paper, it is pointed out that the form of $M_R$ is 
originated in structure of neutrino mass matrix for 
$(\nu_i, N_\alpha)$ where $\nu_i$ ($i=1,2,3$) and $N_\alpha$ 
($\alpha=1,2,3$) are U(3)-family and U(3)$'$-family triplets, 
respectively.
\end{abstract}

PCAC numbers:  
  11.30.Hv, 
  12.60.-i, 
  14.60.Pq,  

\
\vspace{5mm}

{\large\bf 1  \  Introduction} 

Recently, Nishiura and the author \cite{K-N_PRD15R,K-N_MPLA16} 
have proposed a unified mass matrix model under a family symmetry 
U(3)$\times$U(3)$'$:
$$
(\bar{f}_L^i \ \ \bar{F}_L^\alpha ) 
\left(
\begin{array}{cc}
(0)_i^{\ j}  &  \langle\Phi_f\rangle_i^{\ \beta}  \\
\langle \bar{\Phi}_{f} \rangle_\alpha^{\ j} & 
-\langle \hat{S}_f \rangle_\alpha^{\ \beta} 
\end{array} \right) 
 \left(
\begin{array}{c}
f_{Rj} \\
F_{R\beta}
\end{array} 
\right) , \ \ \ \ (f=u, d, e) 
\eqno(1.1)
$$
where $f_i$ ($i=1, 2, 3)$ and $F_\alpha$ ($\alpha=1, 2, 3$) 
are U(3)-family and U(3)$'$-family triplets, respectively, 
so that we obtain a Dirac mass matrix of $f$-sector as follows,  
$$
(\hat{M}_f)_i^{\ j} = \langle \Phi_f \rangle_i^{\ \alpha}
 \langle \hat{S}_f^{-1} \rangle_\alpha^{\ \beta}
\langle \bar{\Phi}_f\rangle_\beta^{\ j} ,
\eqno(1.2)
$$ 
under a seesaw approximation. 
(Hereafter, we denote U(3)-family nonet scalars as 
a notation  $(\hat{A})_i^{\ j}$ and anti-6-plet scalars  as 
a notation $(\bar{A})^{ij}$.)
Here, the VEV matrices $\langle \Phi_f \rangle$ are given by
$$
\begin{array}{l}
\langle\Phi_e \rangle =  m_{0e} {\rm diag}(z_1, z_2, z_3) , \\
\langle\Phi_d \rangle =  m_{0d} {\rm diag}(z_1, z_2, z_3) , \\
\langle\Phi_u \rangle = m_{0u} {\rm diag}(z_1 e^{i\phi_1}, 
z_2 e^{i\phi_2}, z_3 e^{i\phi_3}) .
\end{array}
\eqno(1.3)
$$
Since we assume that the U(3)$'$ symmetry is broken into a discrete 
symmetry S$_3$, the vacuum expectation value (VEV) of $\hat{S}_f$ 
has, in general, to take a VEV form
$$
\langle \hat{S}_f\rangle = m_{0f} \left( {\bf 1} + b_f X_3 \right) ,
\eqno(1.4)
$$
where ${\bf 1}$ and $X_3$ are defined by
$$
{\bf 1} = \left( 
\begin{array}{ccc}
1 & 0 & 0 \\
0 & 1 & 0 \\
0 & 0 & 1 
\end{array} \right) , \ \ \ \ \ 
X_3 = \frac{1}{3} \left( 
\begin{array}{ccc}
1 & 1 & 1 \\
1 & 1 & 1 \\
1 & 1 & 1 
\end{array} \right) ,  
\eqno(1.5)
$$
and $b_f$ are complex parameters.  
We take $b_e$ as $b_e=0$, so that the parameters $z_i$ are 
fixed as 
$$
z_i = \frac{\sqrt{m_{ei}}}{\sqrt{m_{e1} +m_{e2} + m_{e3}}} ,
\eqno(1.6)
$$
where $(m_{e1}, m_{e2}, m_{e3})=(m_e, m_\mu, m_\tau)$. 
We may approximately regard $m_{ei}$ in Eq.(1.6) as the 
observed charged lepton masses $m_{ei}^{obs}$. 
However, note that the vales $m_{ei}$ in Eq.(1.6) are 
not always the eigenvalues $(\hat{M}_e)_i^{\ i}$ ($i=1,2,3$) 
of $\hat{M}_e$ given in Eq.(1.2), and $m_{ei}$ are,   
in general, given by $\hat{M}_i^{\ i} = k_0 m_{ei}$ 
($i=1,2,3$) with an arbitrary family-number-independent 
constant $k_0$.

The model \cite{K-N_PRD15R} can successfully describe 
the observed quark masses and Cabibbo-Maskawa-Kobayashi 
(CKM) \cite{CKM} mixings, 
especially, not only ratios among $m_{ui}=(m_u, m_c, m_t)$ and
among $m_{id}=(m_d, m_s, m_b)$, but also ratios 
$m_{ui}/m_{dj}$ when we take $m_{0u}=m_{0d}$.  
(The quark mass matrix structure has first been proposed 
by Fusaoka and the author \cite{K-F_ZPC96} from 
the phenomenological point of view.) 
  
In the neutrino sector, 
according to the conventional neutrino seesaw model \cite{seesaw}, 
we consider that the Majorana mass matrix of the left-handed 
neutrino is given by 
$$
(M_\nu)_{ij} = (\hat{M}_\nu)_i^{\ k} (M_R^{-1})_{kl} 
(\hat{M}_\nu^{T})^l_{\ j} ,
\eqno(1.7)
$$
under the Majirana mass matrix $M_R$ of the right-handed 
neutrino $\nu_R$ with a large mass scale, 
where $\hat{M}_\nu$ is a Dirac mass matrix of neutrinos defined as
$(\bar{\nu}_L)^i (\hat{M}_\nu)_i^{\ j} (\nu_R)_j$. 
However, in the U(3)$\times$U(3)$'$ model, the structure of $M_R$ 
has been given by a somewhat strange form 
$$
M_R \propto  \Phi_\nu \hat{M}_u + \hat{M}_u^T \Phi_\nu 
+ \xi_R \hat{M}_\nu (\hat{M}_\nu)^T ,
\eqno(1.8)
$$
where
$$
\hat{M}_\nu \equiv \Phi_\nu \bar{\Phi}_\nu , 
\eqno(1.9)
$$
$$
 \Phi_\nu = m_{0\nu} {\rm diag}(z_1, z_2, z_3)  ,
\eqno(1.10)
$$
similar to Eq.(1.3).  
Note that $M_R$ in Eq.(1.8) includes the up-quark mass matrix $\hat{M}_u$. 
When we use the VEV values of $\hat{M}_u$ fitted in the quark sector, 
the neutrino mass matrix $M_R$ is described by only one parameter $\xi_R$, 
and we can obtain excellent fitting \cite{K-N_PRD15R} for 
the observed neutrino masses and 
Pontecorvo-Maki-Nakagawa-Sakata (PMNS) \cite{PMNS} mixings.  
(A form $M_R$ which is related up-quark mass matrix $M_u$ 
has first been proposed by the author \cite{YR-Mu}
in somewhat different context $M_R \propto (M_u^D)^{1/2} M_e^D + \cdots$.)

The model has a reasonable basis 
as far as the quark sector, while the form $M_R$ (1.8) does not have 
a theoretical grand, 
and it was nothing but a phenomenological assumption.   
In this paper, it is pointed out the structure of $M_R$ is 
originated in the structure of neutrino mass matrix for 
$(\nu_i, N_\alpha)$ where $\nu_i$ and $N_\alpha$  are U(3)-family 
and U(3)$'$-family triplets, respectively.

\vspace{5mm}

{\large\bf 2  \  Basic idea} 

Correspondingly to the seesaw mass matrix (1.6), we consider
a seesaw mass matrix
$$
(M_\nu)_{ij} = (\Phi_\nu)_i^{\ \alpha} (M_R^{-1})_{\alpha\beta}
(\Phi_\nu^T)^\beta_{\ j} .
\eqno(2.1)
$$
(Hereafter, in order to make the transformation property 
in U(3) and U(3)$'$ symmetry visual, 
we use symbols $\circ$ and $\bullet$ instead of indexes 
$i, j, \cdots$ and $\alpha, \beta, \cdots$. ) 
From the definition (1.1) of the  Majorana neutrino mass matrix 
$(\bar{M}_R)^{\circ\circ}$, we introduce a Majorana mass matrix,  
$(M_R)^{\bullet\bullet}$, for $(N_R)_\bullet$,  as follows:
$$
\bar{M}_R^{\circ\circ} = (\bar{\Phi}_\nu^T)^\circ_{\ \bullet}
(\bar{M}_R)^{\bullet\bullet} (\bar{\Phi}_\nu)_\bullet^{\ \circ} .
\eqno(2.2)
$$
When we neglect U(3)$\times$U(3)$'$ indexes, 
from the relation (1.8), i.e.
$$
(M_R)^{\circ\circ} = \xi_R (\Phi_\nu)^4 + \left\{\Phi_\nu \hat{M}_u + 
(\hat{M}_u \Phi_\nu)^T \right\} ,
\eqno(2.3)
$$
we can write $(\bar{M}_R)^{\bullet\bullet}$ as follows:
$$
(M_R)^{\bullet\bullet} =  \xi_R (\Phi_\nu)^2 + \left\{\Phi_\nu^{-1} 
\hat{M}_u + (\Phi_\nu^{-1} \hat{M}_u)^T \right\} 
$$
$$
=   \xi_R (\Phi_\nu)^2 +  \left\{ P_u \hat{S}_u^{-1} \bar{P}_u 
\Phi_\nu  + (transposed) \right\}  ,
\eqno(2.4)
$$
where $\Phi_u \propto \Phi_\nu P_u$ and 
$P_u$ is a scalar with VEV values 
$$
P_u = {\rm diag} (e^{i\phi_1}, 
e^{i\phi_2}, e^{i\phi_3}) .
\eqno(2.5)
$$

For example, by considering U(3)$\times$U(3)$'$ transformation, 
we may consider $(M_R)^{\bullet\bullet}$ as 
$$
(M_R)^{\bullet\bullet} = (M_R^{1st})^{\bullet\bullet} +
(M_R^{2nd})^{\bullet\bullet} ,
\eqno(2.6)
$$
$$
(M_R^{1st})^{\bullet\bullet} = (\Phi_\nu^T)^\bullet_{\ \circ} 
(E^{-1})^{\circ\circ} (\Phi_\nu)_\circ^{\ \bullet} ,
\eqno(2.7)
$$
$$
(M_R^{2nd})^{\bullet\bullet} = \left\{ (\bar{E}^T)^{\bullet\circ} 
(P_u)_{\circ\bullet} 
(\hat{S}_u^T)^\bullet_{\ \bullet} (P_u^{-1})^{\bullet\circ}
 (\Phi_\nu)_\circ^{\ \bullet} 
+ (transposed) \right\} ,
\eqno(2.8)
$$
where we have neglected family-independent parameters.

In the next section, we will discuss a model which leads to
the VEV relation (2.6).  
 
\vspace{5mm}

{\large\bf 3  \  Mass matrix of $(\nu_L, \nu_R, N_L, N_R)$}

For convenience, in this section, we neglect family-independent 
parameters. 

The first term (2.7) suggests a seesaw-like scenario.
Therefore, we would like to consider that the second term is also 
derived from a seesaw-like scenario in the neutrino mass matrix for
$(\nu_L, \nu_R, N_L, N_R)$: 
$$
(M_R^{2nd})^{\bullet\bullet} = \left\{ \langle \bar{E}^T
\rangle^{\bullet\circ} \langle \hat{S}_u^{\prime -1} 
\rangle_\circ^{\ \circ}  \langle \Phi_\nu \rangle_\circ^{\ \bullet} 
+ (transposed) \right\} ,
\eqno(3.1)
$$ 
where $\langle \hat{S}_u^\prime \rangle_\circ^{\ \circ} $ should 
be given by
$$
\langle \hat{S}_u^\prime \rangle_\circ^{\ \circ} =  
\langle P_u \rangle_{\circ\bullet} 
\langle \hat{S}_u^T \rangle^\bullet_{\ \bullet} 
\langle P_u^{-1} \rangle^{\bullet\circ} .
\eqno(3.2)
$$

In this model, $(\bar{M}_R^{1st})$ and $(\bar{M}_R^{2nd})$ 
can be understood by seesaw scenarios (2.7) and (3.1), while 
the relation (3.2) cannot be understood by seesaw scenario. 
Therefore, we consider that the form (3.2) is obtained from  
a SUSY vacuum condition $\partial W/\partial \Theta =0$ for
the following superpotential:
$$
W= {\rm Tr} \left[ \left\{ (\hat{S}'_u)_\circ^{\ \circ} (P_u)_{\circ\bullet} + 
(P_u)_{\circ\bullet} (\hat{S}_u^T)^\bullet_{\ \bullet} \right\} 
(\Theta)^{\bullet\circ} \right] + (transposed) ,
\eqno(3.3)
$$
where $\Theta$ is a flavon with $\langle \Theta \rangle =0$. 

The structures $(M_R^{1st})^{\bullet\bullet}$ and 
$(M_R^{2nd})^{\bullet\bullet}$ suggest the following 
mass matrix for $( (\nu_L)_\circ, (\nu_R^c)^\circ, \\
(N_L)_\bullet, (N_R^c)^\bullet)$:
$$
((\bar{\nu}_L)^\circ \ \ (\bar{\nu}_R^c)_\circ \ \ 
(\bar{N}_L)^\bullet \ \ (\bar{N}_R^c)_\bullet )
$$
$$
\times 
\left(
\begin{array}{cccc} 
\langle E \rangle_{\circ\circ} \ & \ 
\langle \hat{S}'_u \rangle_\circ^{\ \circ} \ \ & \  \
\langle P_u \rangle_{\circ\bullet} \ & \ 
\langle \Phi_\nu \rangle_\circ^{\ \bullet} \\[0.02in]  
\langle \hat{S}^{\prime T}_u \rangle^{\circ}_{\ \circ} \ & 
\ (\ \ \ )^{\circ\circ} \ \ & \  \ (\ \ \ )^{\circ}_{\ \bullet} \ & 
\ \langle \bar{E}^T \rangle^{\circ\bullet}  \\[0.05in] 
\langle P_u^T \rangle_{\bullet\circ} \ & \ (\ \ \ )_\bullet^{\ \circ} \ \ &
 \  \ (\ \ \ )_{\bullet\bullet} \ & 
\ \langle \hat{S}_u \rangle_\bullet^{\ \bullet} \\[0.02in]  
\langle \Phi_\nu^T \rangle^\bullet_{\ \circ} \ & 
\ \langle \bar{E} \rangle^{\bullet\circ} \ \ & \  \ 
\langle \hat{S}_u^T \rangle^\bullet_{\ \bullet} \ & 
\ (\ \ \ )^{\bullet\bullet} \\[0.02in] 
 \end{array} \right) \ \ \ \ 
 \left(
 \begin{array}{c}
 (\nu_L^c)^\circ  \\[0.02in] 
 (\nu_R)_\circ   \\[0.05in] 
 %
  %
  (N_L^c)^\bullet   \\[0.02in] 
  (N_R)_\bullet   \\[0.02in] 
  \end{array}
  \right) . 
\eqno(3.4)
$$
Here, $(\ \ \ )^{\bullet\bullet}$ is a room for would-be 
$(\bar{M}_R)^{\bullet\bullet}$. 
Thus, we can assign all scalars (flavons) in this mass matrix 
(3.4) without duplication.

Finally, we would like to comment on $R$ charge assignment.
We adopt $R$ charge assignment for flavons (scalars) $A$ and 
fermions $\psi$ as follows
$$
R(\bar{A})=R(A) , \ \ \ R(\bar{\psi}_{L/R}) \neq R(\psi_{L/R}) .
\eqno(3.5)
$$
For example, in Eq.(3.4), we have defined the flavon $\hat{S}_u$ as  
$(\bar{N}_L)^\bullet (\hat{S}_u)_\bullet^{\ \bullet} 
(N_R)_\bullet$. 
This does not always mean $R(U_L)=R(N_L)$ and 
$R(U_R)=R(N_R)$ where $(U_L, U_R)$ are components of $(F_L, F_R)$ 
in the sector $f=u$. 
It means only 
$$ 
R(N_L) + R(N_R) = R(U_L) + R(U_R) .
\eqno(3.6) 
$$
Thus, we can put the flavon $\hat{S}_u$ on the desirable position 
in the neutrino mass matrix (3.4). 
As we already stressed, it has an important meaning that
we could assign all scalars (flavons) in this mass matrix 
(3.4) without duplication. 
It means that we can uniquely assign those flavons without 
mixing under suitable $R$-charge assignment for $(\nu_{L/R}, N_{L/R})$.  
  
Also, note that, in the mass matrix (3.4), there is no 
$(E)_{\circ\bullet}$ and $(\bar{P}_u)^{\circ\bullet}$ 
in spite of the existence 
 $(\bar{E})^{\circ\bullet}$ and $(P_u)_{\circ\bullet}$.  
This is possible only under the selection rule (3.5). 
For example, note that a conjugate term of the term  
 $(\bar{\nu}_L)^\circ (P_u)_{\circ\bullet} (N_L^c)^\bullet$ 
is not $(\bar{N}_R^c)_\bullet  (\bar{P})^{\bullet\circ} 
(\nu_R)_\circ$, but $(\bar{N}_L^c)_\bullet 
(\bar{P}_u)^{\bullet\circ} (\nu_L)_\circ$.  
Since 
$$
\begin{array}{l}
R((\bar{E})^{\bullet\circ}) = 2 -R((\nu_R)_\circ) 
-R((\bar{N}_R^c)_\bullet), \\
R((\bar{P}_u)^{\bullet\circ}) = 2 -R((\nu_L)_\circ )
-R((\bar{N}_L^c)_\bullet) , 
\end{array}
\eqno(3.7)
$$
if we take 
$$
R((\nu_R)_\circ) +R((\bar{N}_R^c)_\bullet) \neq 
R((\nu_L)_\circ) +R((\bar{N}_L^c)_\bullet) ,
\eqno(3.8)
$$
we can regard $(P_u)_{\circ\bullet}$ and $\bar{E}^{\circ\bullet}$
as separate flavons.

\vspace{5mm}

{\large\bf 4  \ Scales of VEV matrices}

In the recent study \cite{K-N_IJMPA17} in the 
U(3)$\times$U(3)$'$ model, 
it has been concluded that flavon VEVs with 
U(3)$\times$U(3)$'$ indexes $A_{\bullet\bullet}$,
$B_{\bullet\circ}$ and $C_{\circ\circ}$ take the following scales
$$
\langle A_{\bullet\bullet} \rangle \sim \Lambda_1 \sim 3 \times 10^7\, 
{\rm TeV} , \ \ \ 
\langle B_{\bullet\circ} \rangle \sim \Lambda_2 \sim 3 \times 10^4\, 
{\rm TeV} , \ \ \      
\langle C_{\circ\circ} \rangle \sim \Lambda_3 \sim 9 \, {\rm TeV} . 
\eqno(4.1)
$$

In order that the seesaw scenario $M_R^{1st}$, Eq.(2.7), holds, 
the flavon scales have to satisfy the relation 
$$
\langle E_{\circ\circ} \rangle \gg 
\langle (\Phi_\nu)_\circ^{\ \bullet} \rangle .
\eqno(4.2)
$$
As we have proposed in Ref.\cite{K-N_IJMPA17}, we adopt such the 
mechanism $\langle \Phi_\nu \rangle = \xi_\nu \langle \Phi_e \rangle$ 
with $\xi_\nu \ll 1$.

For the seesaw scenario $M_R^{2nd}$, Eq.(2.8), a VEV relation
$$
\langle \bar{E}^{\bullet\circ} \rangle \sim 
\langle (\Phi_\nu)_\circ^{\ \bullet} \rangle \ll 
\langle (\hat{S}'_u)_\circ^{\ \circ} \rangle,
\eqno(4.3)
$$
is required. 
We consider that a scale of the flavon $(\hat{S}'_u)$ 
is an exceptional case against the general rule (4.1)
in spite of its indexes $(\ )_\circ^{\ \circ}$, 
because the VEV relation has to be
$$
\langle (\hat{S}'_u)_\circ^{\ \circ} \rangle \sim 
\langle (\hat{S}_u)_\bullet^{\ \bullet} \rangle  \sim \Lambda_1 .
\eqno(4.4)
$$
from the consistency among the scales in Eq.(3.3).  

Here, we would like to comment on a scale of SU(2)$_L$. 
Flavons $\Phi_f$ and $\hat{S}_F$ given in (1.1) are 
singlets in the vertical symmetry 
SU(3)$_c \times$SU(2)$_L \times$U(1)$_Y$, 
so that they have only indexes of 
horizontal symmetry (family symmetry). 
The mass matrix (1.1) does not correspond to 
the real masses of quarks and leptons, but it 
represents the Yukawa coupling constant. 
Therefore, note that $f_L$ does not mean 
$f_L =(u, d, e)_L$, and that $f_L$ has to be SU(2)$_L$ singlet.  
In Ref.\cite{K-N_IJMPA17}, the fermions $f_L$ have been 
defined as follows:
$$
f_L \equiv (f_u, f_d, f_\nu, f_e)_L \equiv 
\left( \frac{1}{\Lambda_H} H_u^\dagger q_L, 
\frac{1}{\Lambda_H} H_d^\dagger q_L,
\frac{1}{\Lambda_H} H_u^\dagger \ell_L,
\frac{1}{\Lambda_H} H_d^\dagger \ell_L \right)
\eqno(4.5)
$$
where 
$$
q_L = \left(
\begin{array}{c}
u_L \\
d_L 
\end{array} \right) , \ \ \ \ 
\ell_L = \left(
\begin{array}{c}
\nu_L \\
e^-_L 
\end{array} \right) , \ \ \ \ 
H_u = \left(
\begin{array}{c}
H_u^0 \\
H_u^- 
\end{array} \right) , \ \ \ \ 
H_d = \left(
\begin{array}{c}
H_d^+ \\
H_d^0 
\end{array} \right) . 
\eqno(4.6)
$$
Note that $f_L$ is singlet in SU(2)$_L$, but it has 
U(1)$_Y$ charge. 
Therefore, correspondingly, $f_R$, $F_L$ and $F_R$ are 
SU(2)$_L$ singlets, while they have U(1)$_Y$ charge. 
(We consider that $(F_u, F_d)_{L/R}$ have SU(3)$_c$ 
indexes.) 
For example, the selection of $\hat{S}_f$ in 
$\Phi_f \hat{S}_f^{-1} \bar{\Phi}_f$ 
(for example, 
$\hat{S}_u$ in $\Phi_u \hat{S}_f^{-1} \bar{\Phi}_u$) 
is done by $R$ charge, not by flavor symmetry and/or 
U(1)$_Y$ charge.

Our purpose in this paper was to discuss the neutrino mass 
matrix. 
The neutrino Dirac mass matrix $(\hat{M}_{\nu})_\circ^{\ \circ}$ 
given in Eq.(1.7) comes from the term 
$(1/{\Lambda_H}) \bar{\ell} H_u^\dagger = (1/{\Lambda_H})
(\bar{\nu}_L H_u^0 + \bar{e}_L H_u^-)$ with
$\langle H_u^- \rangle =0$, 
not from Eq.(3.4).

\vspace{5mm}

{\large\bf 5 \ Concluding remarks}

Since the previous U(3)$\times$U(3)$'$ model could give successful 
predictions for the observed quark masses and CKM mixings 
under a reasonable theoretical scenario, while the success 
in the neutrino sector was still phenomenological level. 
We have investigated a possible neutrino mass matrix structure 
in context of the U(3)$\times$U(3)$'$ model.  
As we stressed in Sec.3, it is essential that flavons are uniquely 
assigned in the neutrino mass matrix (3.4) without duplication.  
Under suitable $R$ charge assignment, especially under assumption 
$R(\bar{\psi}) \neq R(\psi)$, we can put $\hat{S}_u$, which was 
defined as $\bar{U}_L \hat{S}_u U_R$ in the up-quark sector, 
into the neutrino sector without confusion. 

In conclusion, we have succeeded in giving 
a theoretical basis to the semi-empirical part 
(the structure of the right-handed neutrino mass matrix $M_R$) 
in the previous U(3)$\times$U(3)$'$ model.  
As a result, the U(3)$\times$U(3)$'$ model has been able to 
become more realistic as a unified mass matrix model of 
quarks and neutrinos.

 \newpage
 

\end{document}